\documentclass[prl,twocolumn,aps,showpacs]{revtex4-1} 
\usepackage{amsmath}
\usepackage{amsfonts}
\usepackage{epsfig}
\usepackage{color}
\usepackage{hyperref}
\usepackage{overpic}

\begin{document}

\newcommand{\todo}[1]{{\small{{\color{red}#1}}}}
\newcommand{\di}{\,\mathrm{d}} % for integral
\newcommand{\dd}{\mathrm{d}} % for differential
\newcommand{\Di}{\,\mathcal{D}} % for functional integral
\newcommand{\pt}{\partial}
\newcommand{\pst}{p^{{\textrm{\scriptsize st}}}}
\newcommand{\pch}{p^{{\textrm{\scriptsize ch}}}}
\newcommand{\tp}{\tilde p}
\newcommand{\bp}{\bar p}
\newcommand{\bP}{\bar P}
\newcommand{\ddp}{\dot{p}}
\newcommand{\xc}{x(\cdot)}
\newcommand{\bxc}{\bar x(\cdot)}
\newcommand{\ddx}{\dot{x}}
\newcommand{\bddx}{\dot{\bar x}}
\newcommand{\Dt}{\D t}
\newcommand{\Dx}{\D x}
\newcommand{\uc}{u(\cdot)}
\newcommand{\buc}{\bar u(\cdot)}
\newcommand{\ddu}{\dot{u}}
\newcommand{\Du}{\D u}
\newcommand{\Dr}{\D r}
\newcommand{\tr}{\tilde r}
\renewcommand{\u}{u} % \renewcommand{\u}{w}
\newcommand{\wc}{u(\cdot)} % \newcommand{\wc}{w(\cdot)}
\newcommand{\ddw}{\dot{u}} % \newcommand{\ddw}{\dot{w}}
\newcommand{\trh}{\tilde\r}
\newcommand{\vek}[1]{\boldsymbol{#1}}
\newcommand{\ddS}{\dot{S}}
\newcommand{\DS}{\D S}
\newcommand{\DStot}{\D S_{\mr{tot}}}
\newcommand{\Sm}{S_{\mr{m}}}
\newcommand{\Ds}{\D s}
\newcommand{\DF}{\Delta\FF}
\newcommand{\DI}{\Delta\II}
\newcommand{\Dp}{\Delta\p}
\newcommand{\Df}{D^{\textrm{\scriptsize (1)}}}
\newcommand{\df}{d^{\textrm{\scriptsize (1)}}}
\newcommand{\tDf}{\tilde D^{\textrm{\scriptsize (1)}}}
\newcommand{\tdf}{\tilde d^{\textrm{\scriptsize (1)}}}
\newcommand{\bDf}{\bar D^{\textrm{\scriptsize (1)}}}
\newcommand{\Dfsq}{D^{\textrm{\scriptsize (1)}^2}}
\newcommand{\Dfx}{D^{\textrm{\scriptsize (1)}\prime}}
\newcommand{\Dg}{D^{\textrm{\scriptsize (2)}}}
\newcommand{\dg}{d^{\textrm{\scriptsize (2)}}}
\newcommand{\tDg}{\tilde D^{\textrm{\scriptsize (2)}}}
\newcommand{\bDg}{\bar D^{\textrm{\scriptsize (2)}}}
\newcommand{\Dgsq}{D^{\textrm{\scriptsize (2)}^2}}
\newcommand{\Dgx}{D^{\textrm{\scriptsize (2)}\prime}}
\newcommand{\Dgxx}{D^{\textrm{\scriptsize (2)}\prime\prime}}
\newcommand{\bDgx}{\bar D^{\textrm{\scriptsize (2)}\prime}}
\newcommand{\Dgxsq}{D^{\textrm{\scriptsize (2)}\prime^2}}
\newcommand{\Dk}{D^{\textrm{\scriptsize $(k)$}}}
\newcommand{\tDk}{\tilde D^{\textrm{\scriptsize $(k)$}}}
\newcommand{\Sc}{S_{u_L}}
\newcommand{\ee}[1]{\mathrm{e}^{#1}}
\newcommand{\mr}[1]{\mathrm{#1}}
\newcommand{\const}{\mr{const}}
\newcommand{\bs}[1]{\boldsymbol{#1}}
\newcommand{\HH}{\mathcal{H}}
\renewcommand{\SS}{\mathcal{S}}
\newcommand{\LL}{\mathcal{L}}
\newcommand{\hLL}{\hat{\mathcal{L}}}
\newcommand{\hLLFP}{\hat{\mathcal{L}}_\mathrm{FP}}
\newcommand{\FF}{\mathcal{F}}
\newcommand{\II}{\mathcal{I}}
\newcommand{\OO}[1]{\mathcal{O}(#1)}
\newcommand{\g}{\gamma}
\newcommand{\G}{\Gamma}
\renewcommand{\b}{\beta}
\renewcommand{\a}{\alpha}
\renewcommand{\d}{\delta}
\newcommand{\e}{\varepsilon}
\newcommand{\eps}{\epsilon}
\newcommand{\p}{\varphi}
\renewcommand{\l}{\lambda}
\renewcommand{\L}{L} % \renewcommand{\L}{\varLambda}
\renewcommand{\r}{r} % \renewcommand{\r}{\rho}
\newcommand{\D}{\Delta}
\newcommand{\z}{\zeta}
\newcommand{\sigi}{\sigma_\infty}
\newcommand{\ra}{\rightarrow}
\newcommand{\la}{\leftarrow}
\newcommand{\Ra}{\Rightarrow}
\newcommand{\lolra}{\longleftrightarrow}
\newcommand{\lora}{\longrightarrow}
\newcommand{\loRa}{\Longrightarrow}
\newcommand{\lola}{\longleftarrow}
\newcommand{\lra}{\leftrightarrow}
\newcommand{\Lra}{\Leftrightarrow}
\newcommand{\lla}{\left\langle}
\newcommand{\rra}{\right\rangle}
\newcommand{\meq}{\stackrel{!}{=}}
\newcommand{\lb}{\left|}
\newcommand{\rb}{\right|}
\newcommand{\eq}[2]{\begin{equation} \ensuremath{\label{#1} #2} \end{equation}}
\newcommand{\eqal}[2]{\begin{equation} \label{#1} \ensuremath{\begin{aligned} #2 \end{aligned}} \end{equation}}
\newcommand{\al}[1]{\begin{align} \ensuremath{#1} \end{align}}
\newcommand{\nn}{\nonumber\\}
\newcommand{\nnn}{\nonumber\\\nonumber}
\newcommand{\intev}{\sum\limits_\pm\int\limits_{-\infty}^{\a_\mr{max}}}
\newcommand{\ka}{\kappa_{\!\pm}\!(\a)}
\newcommand{\rha}{\varrho_{\!\pm}\!(\a)}
\newcommand{\trha}{\tilde\varrho_{\!\pm}\!(\a)}
\newcommand{\abs}{\\[5pt]}
\newcommand{\crs}{\addcontentsline{toc}{section}{cursor}}
\newcommand{\figsize}{\footnotesize}
\newcommand{\dint}{\displaystyle\int}
\newcommand{\abbs}[2]{\includegraphics[trim = 0pt 0pt 0pt 10pt, clip, width=#2\textwidth]{/home/statphys/nickelsen/kile/paper12b/figs/abbs_#1.eps}}
\newcommand{\genft}[2]{\includegraphics[trim = 0pt 0pt 0pt 15pt, clip, width=#2\textwidth]{/home/statphys/nickelsen/kile/paper12b/figs/#1.eps}}
\newcommand{\figadd}[3]{\includegraphics[trim = 0pt 0pt 0pt #3pt, clip, width=#2\textwidth]{/home/statphys/nickelsen/kile/paper12b/figs/#1.eps}}
\newcommand{\figaddc}[3]{\begin{center}\includegraphics[trim = 0pt 0pt 0pt #3pt, clip, width=#2\textwidth]{/home/statphys/nickelsen/kile/paper12b/figs/#1.eps}\end{center}}
\newcommand{\pfad}[1]{/home/statphys/nickelsen/kile/paper12b/figs/#1.eps}

\title{Probing small-scale intermittency with a fluctuation theorem} 
\author{Daniel \surname{Nickelsen}}
\email{d.nickelsen@uni-oldenburg.de}
\author{Andreas \surname{Engel}}
\affiliation{Institut f{\"{u}}r Physik, Carl von Ossietzky Universit{\"{a}}t Oldenburg, 26111 Oldenburg, Germany}

\begin{abstract}
We characterize statistical properties of the flow field in developed turbulence using concepts from stochastic thermodynamics. On the basis of data from a free air-jet experiment, we demonstrate how the dynamic fluctuations induced by small-scale intermittency generate analogs of entropy-consuming trajectories with sufficient weight to make fluctuation theorems observable at the macroscopic scale. We propose an integral fluctuation theorem for the entropy production associated with the stochastic evolution of velocity increments along the eddy-hierarchy and demonstrate its extreme sensitivity to the accurate description of the tails of the velocity distributions. 
\end{abstract}


\maketitle

%%%%%%%%%%%%%%%%%%%%%%%%%%%%%%%%%%%%%%%%%%%%%%%%%%%%%%%%%%%%%%%%%%%%%%%%%%%%%%%%%%%%%%%%%%%%%%%%%%%%%%%%%%%%%%%%%%%%%

All processes in nature are bound to produce entropy. This central dogma of macroscopic thermodynamics got substantially qualified in the preceding decade by new insights into the properties of small, strongly fluctuating systems. If entropy {\em consuming} trajectories occur with appreciable probability, thermodynamic {\em inequalities} may be considerably tightened to assume the form of {\em equalities} \cite{Evans_Cohen_Morriss_1993,Jarzynski_1997}. The emerging field of stochastic thermodynamics (for recent reviews see \cite{Seifert_2012,Jarzynski_2011a}) focuses on the full probability distributions of thermodynamic variables like heat, work, and entropy and establishes thermodynamic relations for individual fluctuating histories of the systems under consideration. Most prominent among these relations are the so-called {\em fluctuation theorems} (FTs) quantifying the relative frequency of entropy-consuming as compared to entropy-producing trajectories. Applications of these developments concern free-energy estimates of biopolymers \cite{Pohorille_Jarzynski_Chipot_2010,Palassini_Ritort_2011}, the efficiency of nano-machines \cite{Seifert_2011,Broeck_Kumar_Lindenberg_2012}, and the thermodynamic cost of information processing \cite{Toyabe_Sagawa_Ueda_Muneyuki_Sano_2010,Mandal_Jarzynski_2012}, to name a few.

On the experimental side, most investigations have been done with {\em nanoscopic} setups like in single-molecule manipulations \cite{Collin_Ritort_Jarzynski_Smith_Tinoco_Bustamante2005,Junier_Mossa_Manosas_Ritort2009,Jarzynski_2011b}, colloidal particle dynamics \cite{Wang_Sevick_Mittag_Searles_Evans_2002,Mehl_Lander_Bechinger_Blickle_Seifert_2012,Ciliberto_Joubaud_Petrosyan_2010} or harmonic oscillators \cite{Ciliberto_Joubaud_Petrosyan_2010}. For these systems, typical free-energy differences are of order $k_\mathrm{B} T$ and the ubiquity of {\em thermal} fluctuations ensures the broad distributions of work, heat, and entropy which are indispensable for the application of fluctuation theorems. Increasing the size of these systems to macroscopic orders, the importance of thermal fluctuations fades, entropy-consuming trajectories become exceedingly rare, and the fluctuation theorems degenerate to the inequalities known from traditional thermodynamics. Besides some investigations in granular media \cite{Feitosa_Menon_2004,Naert_2012,Chong_Otsuki_Hayakawa_2010}, rather few examples of {\em macroscopic} systems have been identified which are amenable to an analysis using FTs.

Turbulent flow of liquids and gases is a fascinating phenomenon with many different facets that has been captivating scientists for centuries. Despite its broad range of technical relevance including turbulent drag \cite{Jimenez_2004}, turbulent mixing \cite{Dimotakis_2005}, atmospheric turbulence with implications for climatic models \cite{Arnfield_2003} and the prospects of wind energy \cite{Böttcher_Barth_Peinke_2006, Milan_Wächter_Peinke_2013}, several aspects of turbulent flows are still not fully understood. In particular, the intricate pattern of small-scale flow in developed turbulence with its intermittent change between laminar periods and violent bursts of activity have eluded a thorough theoretical understanding so far.

%%%%%%%%%%%%%%%%%%%%%%%%%%%%%%%%%%%%%%%%%%%%%%%%%%%%%%%%%%%%%%%%%%%%%%%%%%%%%%%%%%%%%%%%%%%%%%

\begin{figure}
		\includegraphics[width=0.45\textwidth]{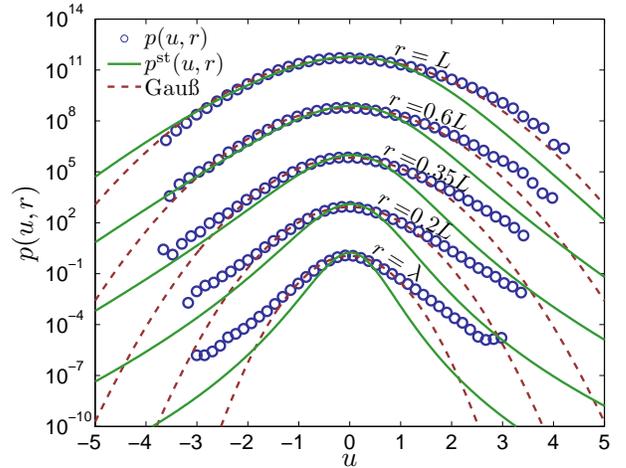}
		\caption{Distribution $p(u,r)$ of velocity increments $u$ at various scales $r$ (circles) in the turbulent flow of a free jet experiment \cite{Renner_Peinke_Friedrich_2001}. The velocity increments $u$ are given in units of the standard deviation  $\sigi=0.54\,\mr{m/s}$ at infinite scales. Also shown is the instantaneous stationary distribution $\pst(u,r)$ defined in (\ref{e_pst}) (full lines) and Gaussian fits to the experimental data (dashed lines). Both the deviation from the Gaussian approximation and from the stationary distribution increases when approaching smaller scales. For the sake of clarity, the distributions for various scales are vertically shifted by $10^3$.}
		\label{f_pdf_ur}
\end{figure}

%%%%%%%%%%%%%%%%%%%%%%%%%%%%%%%%%%%%%%%%%%%%%%%%%%%%%%%%%%%%%%%%%%%%%%%%%%%%%%%%%%%%%%%%%%%%

In the present letter, we show that the fluctuating flow field of developed turbulence 
represents a proper test system for stochastic thermodynamics. The {\em dynamic} fluctuations of turbulence show up at a macroscopic scale and, at the same time, are strong enough to generate ``non-mainstream'' trajectories with sufficient frequency to observe FTs in action. Using data from a free air jet experiment, we elucidate the nature of the entropy-consuming trajectories and demonstrate the convergence of an integrated FT for data sets of rather small size. We further discuss how to use the FT for the statistical description of the flow field.

Applications of FTs to turbulent flow have been discussed before. On the experimental side, fluctuations of the heat flux \cite{Shang_Tong_Xia_2005}, the injected power \cite{Falcon_Aumaître_Falcón_Laroche_Fauve_2008} and the pressure \cite{Ciliberto_Garnier_Hernandez_Lacpatia_Pinton_RuizChavarria_2004}, as well as the motion of tracer particles \cite{Bandi_Cressman_Goldburg_2007} were studied. Numerical investigations concerned fluctuations of the injected power in the shell-model \cite{Aumaître_Fauve_McNamara_Poggi2001,Gilbert_2004} and properties of augmented Navier-Stokes equations in two dimensions \cite{Gallavotti_Rondoni_Segre_2004}. All these investigations focused on variants of the {\em steady-state} FT \cite{Evans_Cohen_Morriss_1993,Gallavotti_Cohen_1995}. The FT we propose in this letter is qualitatively different. It is no steady-state FT but characterizes the {\em non-stationary} stochastic evolution of velocity increments along the eddy-hierarchy. It is somewhat similar in spirit to the detailed FT proposed in \cite{Baiesi_Maes_2005}, which, however, describes the enstrophy cascade in two-dimensional turbulence.  

In a standard setup, isotropic turbulence is generated by injecting energy into the flow by an external force field at a large, so-called integral scale $L$ \cite{Landau_Lifshitz_1987,Frisch_1995}. By repeated break-up of eddies, a self-similar eddy hierarchy forms which is characteristic for developed turbulence \cite{Richardson_1922}. On average, energy is transferred along the cascade to smaller and smaller scales until, due to molecular friction, it is dissipated in the viscous range. The Taylor scale $\l$ marks the length scale above which the influence of dissipation is still negligible. 

A suitable way to characterize the stationary, homogeneous, and isotropic flow field $\vek {v}(\vek{x},t)$ in the {\em inertial range} between $L$ and $\l$ is via the probability density function $p(u,r)$ of longitudinal velocity increments \cite{Frisch_1995}
\begin{equation}
 u(r) := \vek{e}\cdot\big(\vek{v}(\vek{x} + \vek{e}r,t) - 
 \vek{v}(\vek{x},t)\big)\; .
\end{equation}
Here, $r$ denotes the scale at which the velocity difference $u$ is evaluated, $\vek{e}$ is a unit vector and due to the average symmetries of the turbulent flow, the statistical properties of $u$ only depend on $r$. Fig.~\ref{f_pdf_ur} shows histograms of this distribution using data obtained in a turbulent air jet experiment \cite{Renner_Peinke_Friedrich_2001}. In this setup,
$L=6.7\,\mr{cm}$, $\l=6.6\,\mr{mm}$, and the nozzle-based Reynolds number is about $2.7\cdot10^4$. The flow velocity $v(t)$ is measured a distance of 125 nozzle diameters away from the nozzle and then converted to a flow field $v(x)$ by use of the Taylor hypothesis. Chopping $v(x)$ into non-overlapping intervals, $N=5\cdot10^4$ trajectories $\u(r)$ are obtained from which the shown histograms are compiled. As Fig.~\ref{f_pdf_ur} clearly shows, $p(u,r)$ exhibits a Gaussian form for scales $r\approx L$ and develops pronounced non-Gaussian tails towards scales $r\approx\l$. This effect is commonly referred to as {\em small-scale intermittency}, as intermittent bursts in $\vek{v}(\vek{x})$ cause the boosted occurrence of large values of $u$ on small scales \cite{Sreenivasan_Antonia_1997}. 

An inventive approach to characterize the properties of the distribution $p(u,r)$ in the inertial range is to interpret $u(r)$ as realizations of a {\em Markov process} on the eddy hierarchy with the scale $r$ playing the role of time \cite{Friedrich_Peinke_1997a}. The evolution of $p(u,r)$ is then described by a Master equation with initial condition at $r=L$, for which a Kramers-Moyal expansion \cite{Risken_1989} may be performed. For a variety of experimental situations, the Markovian character of $u(r)$ was verified, and the coefficients $\Dk$ of the corresponding Kramers-Moyal expansion were determined on the basis of experimental data \cite{Marcq_Naert_1998,Friedrich_Peinke_Sahimi_Tabar_2011,Honisch_Friedrich_2011,Kleinhans_2012}. Moreover, in the limit of large Reynolds number, it is possible to systematically derive the Master equation governing the evolution of $p(u,r)$ from the underlying Navier-Stokes equations of the fluid flow and to recursively calculate the coefficients $\Dk$ \cite{Yakhot_1998,Davoudi_Tabar_1999}. In either way, one finds that drift and diffusion coefficients, $\Df$ and $\Dg$ respectively, have well-defined, non-zero limits, whereas all higher coefficients in the Kramers-Moyal expansion vanish asymptotically. We are thus left with a Fokker-Planck equation (FPE) of the form 
\begin{equation} \label{e_FPE_cond_ito}
	\begin{split}
		-\pt_r p(u,r|u_L,L) = \big[
			&-\pt_u \Df(u,r) \\ 
			&+ \pt_u^2 \Dg(u,r)\big] p(u,r|u_L,L) 
	\end{split}
\end{equation}
ruling the statistics of velocity increments on the eddy-hierarchy of developed turbulence. The minus sign on the l.h.s. of the FPE indicates that the evolution proceeds from large to small scales.  

The drift and diffusion coefficients, $\Df$ and $\Dg$, typically depend on $r$ and $u$; for the data shown in Fig.~\ref{f_pdf_ur} one obtains, e.g., \cite{Renner_Peinke_Friedrich_2001}
\begin{align}
 \Df(u,r) &= -a_0r^{0.6} - a_1r^{-0.67}u + a_2u^2 - a_3r^{0.3}u^3 \label{e_D1} \\
 \Dg(u,r) &=  b_0r^{0.25} - b_1r^{0.2}u + b_2r^{-0.73}u^2 \label{e_D2}
\end{align}
with
\begin{align}
 &a_0 = 0.0015 ,\; a_1 = 0.61   ,\; a_2 = 0.0096 ,\; a_3 = 0.0023 \; , \nnn
 &b_0 = 0.033 ,\; b_1 = 0.009 ,\; b_2 = 0.043 \; .
\end{align}
The stochastic dynamics defined by \eqref{e_FPE_cond_ito} therefore exhibits characteristics of a {\em driven non-equilibrium} system. This is apparent also from the difference between $p(u,r)$ and the {\em instantaneous} stationary distribution of the FPE (\ref{e_FPE_cond_ito}) for fixed scale $r$ given by  
\begin{equation} \label{e_pst}
	\pst(u,r) = \frac{\ee{-\p(u,r)}}{Z(r)} \;,\quad 
 Z(r)=\int \ee{-\p(u,r)} \di u
\end{equation}
with the stochastic potential 
\begin{equation}\label{e_defphi}
 \p(u,r) = \ln\Dg(u,r)-\int\limits_{-\infty}^{u} \frac{\Df(u',r)}{\Dg(u',r)} \di u' \; .
\end{equation} 
Examples of $\pst(u,r)$ have been included into Fig.~\ref{f_pdf_ur}. 

In the spirit of stochastic thermodynamics \cite{Seifert_2012}, we now associate with every individual trajectory $\u(r)$ a total {\em entropy production}
\begin{equation} \label{e_DStot_secondlaw}
	\begin{split}
	  \DStot[u(\cdot)] = &-\int\limits_{L}^{\l} \partial_r u(r)\,\partial_u \p\big(u(r),r\big) \di r \\
			&-\ln\frac{p(u_\l,\l)}{p(u_L,L)} .
	\end{split}
\end{equation}
In the usual thermodynamic setting, the first term on the r.h.s. of \eqref{e_DStot_secondlaw} would describe the heat exchange with the reservoir, whereas the second one gives the entropy change of the system itself. The total entropy production \eqref{e_DStot_secondlaw} fulfills the integral FT  \cite{Seifert_2005}
\begin{equation} \label{e_FT}
  \lla \ee{-\DStot} \rra_{\uc} = 1 \; ,
\end{equation}
where the average is over the different realizations of $u(r)$.  

\begin{figure}
		\begin{overpic}[trim = 0pt 0pt 0pt 0pt, clip, width=0.45\textwidth]{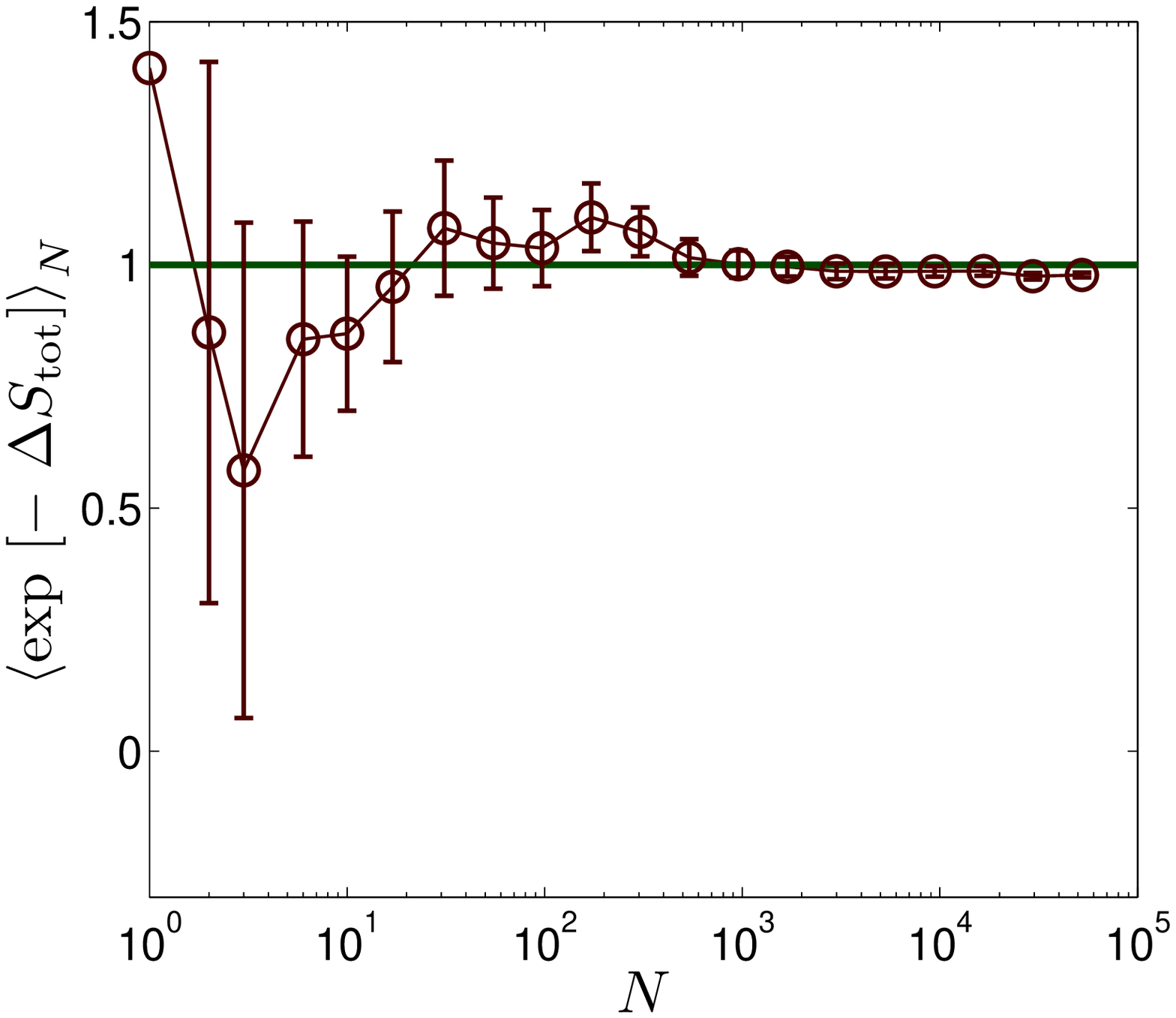}
		\put(30,12){ 
			\includegraphics[trim = 0pt 0pt 0pt 0pt, clip, width=0.29\textwidth]{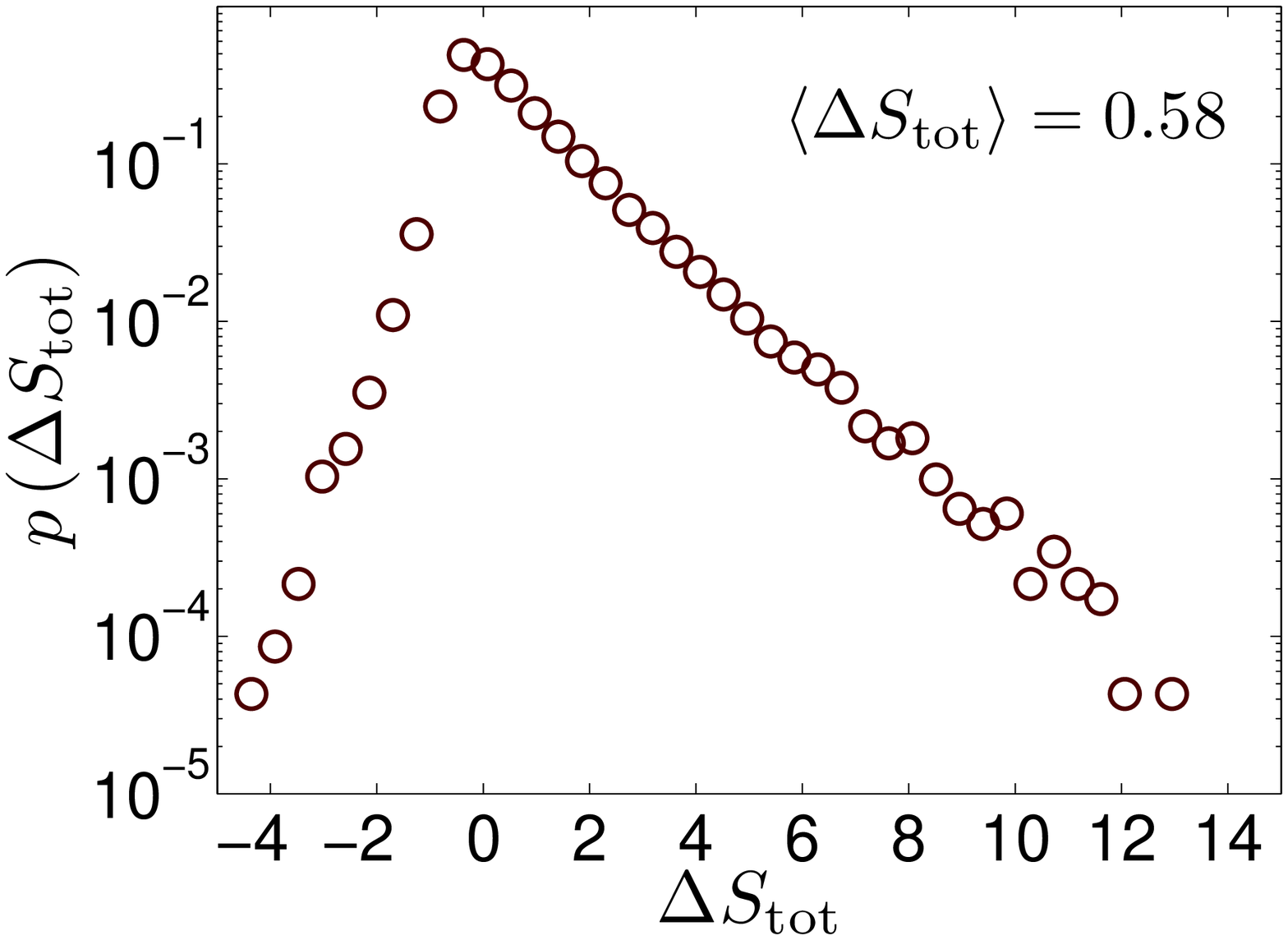} }
  \end{overpic}
		\caption{Empirical average $\lla\exp-\DStot\rra_N$ defined in \eqref{e_emav} for the experimental data of Fig.~\ref{f_pdf_ur} as a function of the sample size $N$. According to the fluctuation theorem (\ref{e_FT}), the average has to converge to the horizontal line. The inset depicts the corresponding distribution of the total entropy production $\DStot$ as defined by (\ref{e_DStot_secondlaw}).}
		\label{f_jfm_exp_ft}
\end{figure}

A reliable estimate of the exponential average in (\ref{e_FT}) on the basis of a finite sample set is possible only if trajectories with $\DStot[\uc]<0$ occur with sufficient frequency. We have used subsets of size $N$ of the realizations for $u(r)$ underlying Fig.~\ref{f_pdf_ur} together with their entropy productions determined by \eqref{e_DStot_secondlaw} and calculated the empirical average 
\begin{equation}\label{e_emav}
 \lla \ee{-\DStot} \rra_N =\frac{1}{N}\sum_{i=1}^N \ee{-\DStot^{(i)}}
\end{equation} 
corresponding to \eqref{e_FT}. The results shown in Fig.~\ref{f_jfm_exp_ft} demonstrate that convergence to the asymptotic value is rather fast. This is corroborated by the appreciable weight of trajectories with {\em negative} entropy production in the distribution $p(\DStot)$ shown in the inset. The macroscopic fluctuating flow fields of developed turbulence therefore share important features with the thermodynamic variables of nanoscopic non-equilibrium systems under the influence of thermal noise. In particular, in both cases the respective probability distributions are sufficiently broad to allow an application of the concepts of stochastic thermodynamics.

\begin{figure}
	\includegraphics[trim = 0pt 0pt 0pt 0pt, clip, width=0.45\textwidth]{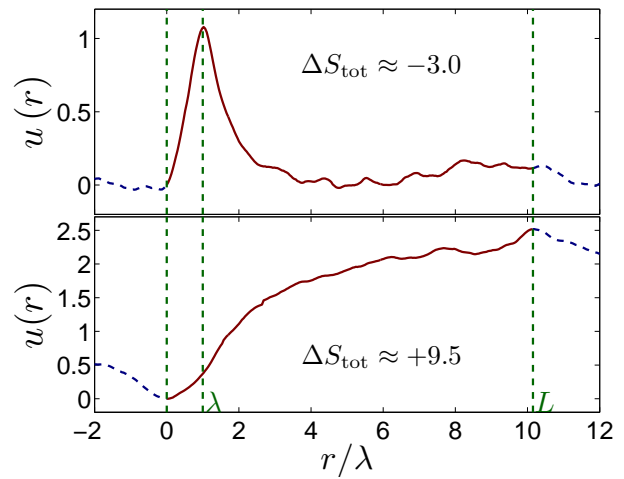}
	\caption{Typical form of measured velocity increments $u(r)$ (full lines) realizing a very small (top) and a very large (bottom) entropy production $\DStot$ defined by (\ref{e_DStot_secondlaw}). The dashed lines show the average part of $v(x)$ neighboring $\uc$. The Taylor scale $\l$ and integral scale $L$ are indicated by vertical lines.}
		\label{f_exp_rare-v}
\end{figure}

The convergence of the empirical average \eqref{e_emav} to the theoretical value 1 given by \eqref{e_FT} also indicates that the drift and diffusion coefficients \eqref{e_D1}, \eqref{e_D2}, estimated on the basis of the experimental data, describe the stochastic properties of the process $u(r)$ rather well. Conversely, by monitoring \eqref{e_emav} during the numerical estimation of $\Df$ and $\Dg$, one has a simple, ``on-the-fly'' criterion to quantify the accuracy of this estimation with an emphasis on the precise modeling of entropy-consuming events. The method presently used for the verification of $\Df$ and $\Dg$ involves the numerical solution of the FPE with the estimated drift and diffusion coefficients and a comparison with the underlying experimental trajectories \cite{Renner_Peinke_Friedrich_2001}, which is, of course, a much more cumbersome procedure.

It is interesting to elucidate some characteristics of the entropy-consuming trajectories. To contrast entropy-consumption with entropy production, we show in Fig.~\ref{f_exp_rare-v} the {\em average} of 50 extreme sequences $u(r)$ giving rise to very small and very large values of $\DStot$ respectively. These averages display the distinct features common to all individual trajectories of the corresponding class. As expected, trajectories giving large and small values of $\DStot$ look rather different from each other. Large entropy production, as shown in the bottom panel of Fig.~\ref{f_exp_rare-v}, is related to a continuous decrease of $u$ for decreasing $r$. In contrast, {\em negative} values of $\DStot$ require violent fluctuations at {\em small} scales together with a smooth flow at large scales as shown in the top panel of Fig.~\ref{f_exp_rare-v}. Therefore, the same class of fluctuations that causes small-scale intermittency in developed turbulence also ensures the good convergence of the integral FT \eqref{e_FT}.

This connection becomes also apparent when studying the deviations from dimensional scaling in developed turbulence. Consider the moments 
\begin{equation} \label{e_Sn_Def}
  S^n(r) = \int u^n\,p(u,r)\di u
\end{equation}
of the distribution $p(u,r)$. The self-similar eddy-hierarchy suggests scaling laws for these moments of the form $S^n(r)\propto r^{\z_n}$ defining the scaling exponents $\z_n$. A relation for these exponents, the so-called {\em K62 scaling}, was proposed in 1962 by Kolmogorov and Oboukhov on the basis of dimensional analysis and some simplifying assumptions about the stochastic energy transfer between scales \cite{Kolmogorov_1962,Oboukhov_1962}:
\begin{equation} \label{e_Sn_K62}
  \z_n = \frac{n}{3} - \mu\frac{n(n-3)}{18} \; .
\end{equation}
The intermittency factor $\mu$ describes deviations from pure dimensional (K41  \cite{Kolmogorov_1941}) scaling. It is an experimental fit factor with typical values of about $0.25$  \cite{Frisch_1995}. For the data of Fig.~\ref{f_pdf_ur}  $\mu\approx0.227$. 

Choosing  
\begin{equation} \label{e_DuD_K62}
	\Df(u,r) = -\frac{3+\mu}{9r}u \;,\qquad \Dg(u,r) = \frac{\mu}{18r}u^2 \;,
\end{equation}
the stochastic dynamics \eqref{e_FPE_cond_ito} reproduces the K62 scaling (\ref{e_Sn_K62}) for the moments \eqref{e_Sn_Def} for any initial distribution $p(u_L,L)$ \cite{Renner_Peinke_Friedrich_2001,Friedrich_Peinke_Sahimi_Tabar_2011}. Note that this is already the most general case: In order to find a scaling law $S^n(r)\propto r^{\z_n}$ from the Fokker-Planck dynamics \eqref{e_FPE_cond_ito}, one must have $\Df\sim u/r$ and $\Dg\sim u^2/r$ \cite{Hosokawa_2002}. 

These dependencies are, however, also special with respect to the FT \eqref{e_FT}. Given \eqref{e_DuD_K62}, we may transform to logarithmic ``time'' $\log L/r$ to end up with a FPE describing a {\em stationary} process without external driving. The FT then merely describes the {\em relaxation} process from an initial non-equilibrium distribution to the stationary state $\pst=\d(u)$ where all $S^n(r)\to0$ \cite{Friedrich_Peinke_1997a,Esposito_Broeck_2010a}. Corrections to K62 scaling therefore correspond to a non-trivial ``time'' dependence of drift and diffusion coefficients in the FPE and hence express genuine non-equilibrium dynamics along the eddy-hierarchy. 

To highlight the sensitivity of the FT to small-scale intermittency, we 
specify \eqref{e_FT} to the drift and diffusion coefficients \eqref{e_DuD_K62} of K62 scaling. Using \eqref{e_defphi} and \eqref{e_DStot_secondlaw}, we find 
\begin{equation} \label{e_FT_K62}
  \lla \frac{u_r^\nu \, p_r(u_r)}{u_L^\nu \, p_L(u_L)} \rra = 1\; ,
\end{equation}
with $\nu=\frac{6+4\mu}{\mu}\approx28$. This large value of $\nu$ is consistent with the qualitative picture discussed above: Trajectories corresponding to large values of $\DStot$ have $u_L>u_r$, whereas those with negative $\DStot$ feature $u_L<u_r$. Using data from numerical simulations of the Langevin equation corresponding to \eqref{e_DuD_K62}, we indeed find a smooth convergence of \eqref{e_FT_K62} for sample sizes of $10^4$ or larger.

The crucial point, however, is that \eqref{e_FT_K62} fails dramatically for realistic turbulent flows. Using again the experimental data of Fig.~\ref{f_pdf_ur}, the average in \eqref{e_FT_K62} results into about $10^{70}$ instead of $1$! The value 1 is only approached if small-scale fluctuations occur with the frequency characteristic for the K62 model. The much more frequent and stronger fluctuations of a {\em realistic} turbulent flow, however, cause the rapid divergence of (\ref{e_FT_K62}), which we explain by the well known fact that K62 underestimates the frequency of large fluctuations on small scales (i.e., the scaling \eqref{e_Sn_K62} is only good for $n\lesssim10$ \cite{Anselmet_Gagne_Hopfinger_Antonia_1984,Frisch_1995}). Hence, the corresponding failure of K62 to accurately describe the {\em tails} of $p(u,r)$ is most strikingly demonstrated by the breakdown of \eqref{e_FT_K62}.

In conclusion, we have shown that the violent small-scale fluctuations in turbulent flows make developed turbulence an interesting model system for stochastic thermodynamics. We have proposed an integral fluctuation theorem that characterizes the stochastic evolution of velocity increments along the eddy-hierarchy which is extremely sensitive to the precise modeling of small-scale intermittency. Moreover, it may be used as a simple ``sum-rule'' to quantify the accuracy of parameter estimation from experimental data drawn from turbulent flows. As also other models of developed turbulence like those yielding  
scaling laws different from K62 \cite{She_Leveque_1994,Lvov_Procaccia_2000}, propagator methods \cite{Castaing_Gagne_Hopfinger_1990}, or field-theoretic approaches \cite{Yakhot_1998,Yakhot_2003} correspond to a Markovian dynamics of velocity increments on the eddy-hierarchy \cite{Hosokawa_2002,Amblard_Brossier_1999,Davoudi_Tabar_1999} it should be interesting to apply our analysis also to these approaches.

% acknowledgments
{\em Acknowledgments.} We thank C. Renner, R. Friedrich and J. Peinke for providing us with excellent data, and R. Friedrich, C. Honisch, O. Kamps, M. Niemann, J. Peinke and M. W\"achter for fruitful discussions. DN acknowledges financial support from the Deutsche Forschungsgemeinschaft under grant EN 278/7 and from the Heinz-Neum\"uller foundation.\\

% references
\bibliographystyle{h-physrev}
\bibliography{allrefs}

\end{document}